\documentclass[a4paper,11pt]{article}
\pdfoutput=1
\usepackage{jcappub}
\usepackage{graphicx,%
bm,%
multirow,%
array,%
cancel,%
booktabs,%
mathtools,%
slashed,%
amsmath,%
amsfonts,%
amssymb,%
stmaryrd,%
overpic,%
rotating,%
subcaption
}
\usepackage[utf8]{inputenc}
\usepackage[english]{babel}
\usepackage[usenames,dvipsnames,svgnames]{xcolor}
\usepackage[T1]{fontenc}
\graphicspath{{./}{./Figures/}}

\def\mn{{\mu\nu}}
\def\H{\mathcal{H}}
\def\so{\quad\Rightarrow\quad}
\def\dd{{\rm d}}

\renewcommand{\tilde}{\widetilde}
\allowbreak
\allowdisplaybreaks

\title{\texttt{fRevolution} -- Relativistic Cosmological Simulations in \texorpdfstring{$f(R)$}{f(R)} Gravity I: Methodology}
\author[a]{Lorenzo Reverberi}
\author[b]{and David Daverio}

\affiliation[a]{CEICO, Institute of Physics of the Czech Academy of Sciences, Na Slovance 2, 18221 Praha 8, Czech Republic}

\affiliation[b]{Centre for Theoretical Cosmology, Department of Applied Mathematics and Theoretical Physics, Wilberforce Road, Cambridge CB3 0WA, United Kingdom}

\emailAdd{lorenzo.reverberi@fzu.cz}
\emailAdd{dd415@cam.ac.uk}

\abstract{
We present the new relativistic cosmological particle-mesh code \texttt{fRevolution}, based on~\texttt{gevolution}~\cite{Adamek:2016zes}, aimed at simulating non-linear structure formation in $f(R)$ gravity. We introduce the general framework and approximation scheme, and the set of equations used to solve for the full set of gravitational perturbations. We show results for a point mass field and for cosmological simulations in the Hu-Sawicki model, and compare them to those of existing Newtonian codes. A more detailed analysis and discussion of our solutions will be carried out in a following paper~\cite{Reverberi_Davero_et_al_preparation}.
}

\begin{document}
\maketitle
\flushbottom

\section{Introduction}

The current and upcoming large scale structure surveys will be able to test cosmological structure formation with unprecedented precision. To properly understand and interpret this data, we therefore need to increase the precision of our theoretical prediction as well. The standard approach to investigate the non-linear process of structure formation is N-body simulations~\cite{Teyssier:2001cp,Springel:2005mi,Potter:2016ttn}, which however ignore any relativistic effect. Nevertheless, recent works have shown that it is possible to overcome this intrinsic limitation by interpreting the predictions of a Newtonian simulation in a relativistic context~\cite{Chisari:2011iq,Green:2011wc,Fidler:2015npa,Fidler:2016tir,Borzyszkowski:2017ayl,Fidler:2017pnb}. Such an approach is well defined in the weak field approximation~\cite{Green:2011wc,Green:2010qy}, and allows to rely on Newtonian N-body codes to predict the matter dynamics of a relativistic theory such as General Relativity (GR).
It is worth to note that it is possible to include relativistic effect in the initial conditions \cite{Fidler:2017ebh, Adamek:2017grt}, include relativistic species~\cite{Brandbyge:2016raj, Fidler:2018bkg} and, finally, to reconstruct relativistic observables~\cite{Fidler:2017pnb}.

On the other hand, in the Newtonian framework, the scale factor is completely decoupled from the evolution of matter and therefore needs to be set by hand. Therefore when one wants to implement dynamical dark energy or modified gravity theories, even if such theory aims to modify the dynamic of the scale factor, the latter is \textit{de facto} reconstructed and set by hand, or taken to be the one of the $\Lambda$--Cold Dark Matter concordance model ($\Lambda$CDM)~\cite{Oyaizu:2008sr, Oyaizu:2008tb, Schmidt:2008tn, Li:2011vk, Li:2012by, Puchwein:2013lza, Brax:2012nk, Brax:2013mua, Llinares:2013jza}. While this is justified in first approximation for most viable alternative models, we cannot expect to be able to capture all the new dynamics following this approach, which thus might limit our ability to constrain these theories.

This motivates the creation of the code~\texttt{gevolution}, based on GR and directly based on the weak field approximation~\cite{Adamek:2013wja,Adamek:2015eda,Adamek:2016zes}. In this paper, we propose a method to extend it to $f(R)$ gravity models and discuss its implementation in the new code \texttt{fRevolution} and the first results.

\subsection{Metric}
\label{sec_basics}

We consider the line element of a perturbed FLRW metric in the Poisson gauge
\begin{equation}
	\label{eq_line_element}
	\dd s^2 = a^2(\tau)\left[-(1+2\Psi)\dd \tau^2 - 2B_i \dd x^i \dd \tau + (1-2\Phi) \delta_{ij} \dd x^i \dd x^j + h_{ij}\dd x^i \dd x^j \right].
\end{equation}
where $a(\tau)$ is the cosmological scale factor, $\tau$ is the conformal time, and $x^i$ are the comoving Cartesian coordinates. As usual, Greek indices run through all spacetime dimensions, while Latin indices run only on space-like dimensions. An overdot will denote derivative with respect to $\tau$, a bar will denote background quantities and a tilde will denote Fourier transforms. The Hubble parameter is defined as $\H \equiv \dot a/a$.

The Poisson gauge corresponds to choosing $B_i$ divergenceless, and $h_{ij}$ traceless and divergenceless\footnote{We should point out that in the present version of the code we neglect the tensor perturbations $h_{ij}$ and their back-reaction on the particle evolution. Firstly, the effect of $h_{ij}$ is much smaller than that of $B_i$ which is itself much smaller than that of the standard scalar potentials $\Phi$ and $\Psi$. Secondly, resolving the full dynamics of massless tensor perturbations is computationally extremely expensive. One can still recover the approximate but pretty accurate configuration of $h_{ij}$ at each time by neglecting the time derivatives. See the original \texttt{gevolution} paper~\cite{Adamek:2016zes} for details on this point.}, that is
\begin{equation}
	\delta^{ij}\partial_i B_j = 0,\quad \delta^{ij}h_{ij} = \delta^{ij}\partial_i h_{jk} = 0\,.
\end{equation}
We can maintain these even beyond the linear level provided that we remain in the weak field regime, where all the metric perturbations remain small $\ll 1$. Following the \texttt{gevolution} prescription, we also define the gravitational slip
\begin{equation}
	\chi \equiv \Phi - \Psi\,,
\end{equation}
and where convenient we will use the auxiliary field
\begin{equation}
	\mathcal B_i \equiv a^{-2}B_i\,.
\end{equation}

\subsection{\texorpdfstring{$f(R)$}{f(R)} Gravity}
\label{sec_Hu_Sawicki}
Our goal is to study structure formation in $f(R)$ gravity. Among the possible modified gravity alternatives for cosmic acceleration~\cite{Clifton:2011jh, Tsujikawa:2010zza, Koyama:2015vza}, $f(R)$ gravity is one of the more popular and well-studied classes of theories. There is an extensive literature on $f(R)$ gravity and its cosmological implications, for a review see for instance~\cite{Sotiriou:2008rp, DeFelice:2010aj, Capozziello:2011et} and references therein.

The action of the theory is\footnote{We choose to work in the Jordan frame, where matter is minimally coupled to gravity so that the geodesics are the same as in GR. Alternatively, one can perform a conformal transformation to the Einstein frame, where the gravitational action is Einstein-Hilbert $S \sim \int \dd^4 x\,\tilde R(\tilde g_\mn)$, but matter is coupled to a different metric than $\tilde g_\mn$, so that the additional complexity resides in computing non-standard geodesics instead of non-standard evolution for the metric perturbations. Both approaches are equally valid and must lead to the same observable predictions.}
\begin{equation}
	\label{eq_fR_action}
	S = \frac{1}{16\pi G}\int d^4x\,\sqrt{-g}\,F(R) \equiv \frac{1}{16\pi G}\int d^4x\,\sqrt{-g}\,[R + f(R)] + S_m[\psi; g_\mn].
\end{equation}
where $f$ is a non-linear function of the Ricci scalar $R$. The field equations read
\begin{equation}
	(1 + f_R) R_\mn - \frac{R + f}{2}\,g_\mn + \square_\mn f_R = 8\pi G T_\mn\,,
\end{equation}
where we denoted $\square_\mn \equiv g_\mn\square - \nabla_\mu\nabla_\nu$ for compactness.

Among the $f(R)$ models relevant for the cosmic acceleration, the Hu-Sawicki model~\cite{Hu:2007nk} is likely the most studied and tested at the level of cosmological background, linear perturbations, and Solar System (post-Newtonian approximation)~\cite{Hu:2007pj, Pogosian:2007sw, Capozziello:2007eu, Martinelli:2009ek, Cataneo:2014kaa, Koyama:2015vza}, and at the non-linear level with the use of simulations~\cite{Oyaizu:2008sr, Oyaizu:2008tb, Schmidt:2008tn, Li:2012by}. These studies include possible degeneracies with other effects produced, for instance, by massive neutrinos~\cite{Baldi:2013iza}. The model is given by
\begin{equation}
	f(R) = -m^2 \frac{c_1 (R/m^2)^n}{1 + c_2 (R/m^2)^n}\,,
\end{equation}
where typically $m^2$ is of the order of the present curvature of the Universe. At large curvatures, this can be approximated by
\begin{equation}
	\begin{aligned}
		f &\approx -m^2\frac{c_1}{c_2} + m^2 \frac{c_1}{c_2^2}\left(\frac{m^2}{R}\right)^n\,,\qquad
		f_R &\approx -\frac{n\,c_1}{c_2^2}\left(\frac{m^2}{R}\right)^{n+1}
	\end{aligned}
\end{equation}
Moreover, in the limit $c_1/c_2^2 \to 0$ at fixed $c_1/c_2$ the solutions are approximately~\cite{Hu:2007nk}
\begin{equation}
	R \approx 8\pi G \rho_m - 2f \approx 8\pi G \rho_m + 2m^2\frac{c_1}{c_2}\,,
\end{equation}
so choosing
\begin{equation}
	\label{eq_HS_c1c2}
	\frac{c_1}{c_2} = \frac{16\pi G \rho_0 \Omega_\Lambda}{m^2}
\end{equation}
will produce the observed accelerated background expansion. In most situations, the single parameter
\begin{equation}
 f_{R_0} \equiv f_R(R_0) \approx - \frac{n\,c_1}{c_2^2}\left(\frac{m^2}{R_0}\right)^{n+1}
\end{equation}
where $R_0 = 8\pi G \rho_0 (\Omega_m + 4\Omega_\Lambda)$ is the present cosmological curvature, is enough to essentially specify all the model dynamics, and it is generically found that for small enough values of $|f_{R_0}| \lesssim 10^{-5}$, the Hu-Sawicki model is in agreement with all existing observations\footnote{Possible issues related to the quasi-static approximation and to past singularities~\cite{Frolov:2008uf, Reverberi:2012ew, Noller:2013wca, Llinares:2013jua, Sawicki:2015zya} suggest that we might be able to put even stronger constraints once these effects are taken into account; however, these issues are beyond the scope of this work and we will not consider them further.}. Operationally, specifying $f_{R_0}$, $m^2$ and $n$ completely determines $c_1$ and $c_2$ once we impose the condition~\eqref{eq_HS_c1c2}. In our simulations, we take $n=1$ and consider three values of $|f_{R_0}| = 10^{-4}, 10^{-5}, 10^{-6}$ (see later for more details).

\section{Cosmological Background}
\label{sec_fR_cosmo_background}

Because we work in the Jordan frame, the matter evolution of the different species remains the same as in GR, so that we only need the scale factor and the present (or initial) abundances in order to compute the background density and pressure:
\begin{equation}
	\bar\rho = \bar T^0_0 = \sum_i \Omega_i a^{-3(1 + w_i)}\qquad \bar P \delta^j_k = \bar T^j_k = \delta^j_k\sum_i \Omega_i w_i a^{-3(1+w_i)}
\end{equation}
where as usual
\begin{equation}
	w_i =
	\begin{cases}
		0 & \text{non relativistic matter: baryons, CDM}\\
		1/3 & \text{relativistic species, radiation}\\
		-1 & \text{cosmological constant}
	\end{cases}
\end{equation}
The background equations are given by
\begin{equation}
	\label{eq_fR_background_new}
	(1+ \bar f_R)\bar R_\mn + \frac{\bar R + \bar f}{2}\bar g_\mn + \bar \square_\mn \bar f_R = 8\pi G \bar T_\mn\,,
\end{equation}
and their trace is
\begin{equation}
	\label{eq_fR_background_trace}
	3\ddot{\bar f}_R + 6\H\dot{\bar f}_R + \left(2\bar f + \bar R - \bar f_R \bar R \right)a^2 = -8\pi G a^2 \bar T^\mu_\mu\,,
\end{equation}
which we can rewrite as
\begin{equation}
	\label{eq_fR_background_trace_code}
	3 \bar f_{RR}\ddot{\bar R} + 3 \bar f_{RRR}\dot{\bar R}^2 + 6\H \bar f_{RR}\dot{\bar R} + (2\bar f + \bar R - \bar f_R \bar R)a^2 = -8\pi G a^2 \bar T^\mu_\mu\,.
\end{equation}
Although for $f(R)$ models of cosmic acceleration such as Hu-Sawicki the background is essentially that of $\Lambda$CDM at large curvatures, this is not necessarily true at very late times and surely not necessarily true for a generic $f(R)$ model. To facilitate extending our framework to other models and parameters, we decided to keep the background solver completely general. Moreover, because we are interested in looking for large scale relativistic effects, like back-reaction, we should make sure that the correct background is subtracted when calculating the perturbation equations lest we introduce unphysical homogeneous modes in the perturbations. Notice in particular that in general we will \textit{not} have
\begin{equation}
	\bar R = - 8\pi G \bar T\,,
\end{equation}
because of oscillatory solutions and/or of $\Lambda$-like components in the $f(R)$ solutions, which are not present explicitly in the matter energy-momentum tensor. In the code, we solve~\eqref{eq_fR_background_trace_code} using a Runge-Kutta-Fehlberg method~\cite{Fehlberg1970} starting from the ``GR'' initial conditions
\begin{equation}
	\begin{aligned}
		\H_{\rm in} &= \frac{8\pi G}{3}\left(\Omega_m a_{\rm in}^{-3} + \Omega_r a_{\rm in}^{-4} + \Omega_\Lambda\right)\\
		R_{\rm in} &= 8\pi G \left(\Omega_m a_{\rm in}^{-3} + 4 \Omega_\Lambda\right) \\
		\dot R_{\rm in} &= -24\pi G \H_{\rm in}\Omega_m a_{\rm in}^{-3}
	\end{aligned}
\end{equation}
deep in the matter-domination era, typically at redshift $z_{\rm in} = (1 + a_{\rm in})^{-1} \sim 100$.

In order to resolve the curvature and Hubble oscillations, one must choose a time step smaller than the typical oscillation time, which is roughly
\begin{equation}
	\tau_{\rm osc}^2 \simeq a\, \bar f_{RR}^{-1}\,.
\end{equation}

\section{Perturbations}
\label{sec_perturbations}

In order to remove the background we expand each quantity $Q$ as
\begin{equation}
	Q = \bar Q + \delta Q\,,
\end{equation}
keeping in mind that the background quantities of $f$ and its derivatives are computed at $R=\bar R$. For convenience, we define the \textit{scalaron} field
\begin{equation}\label{eq_def_xi}
	\delta f_R \equiv \delta f_R = f_R - \bar f_R \,.
\end{equation}
The perturbation equations then read
\begin{equation}
	\begin{aligned}
		& (1+\bar f_R)\delta R_\mn + \bar R_\mn \delta f_R - \frac{\bar R + \bar f}{2}\delta g_\mn - \frac{\delta R + \delta f}{2}(\bar g_\mn + \delta g_\mn) + \\
		& \hspace{.36\textwidth} + (\delta \square_\mn)f_R + \bar\square_\mn \delta f_R = 8\pi G \delta T_\mn
	\end{aligned}
\end{equation}
So far these are fully general, and do not assume small perturbations. Naturally, we do expect some perturbations to be indeed small, which simplifies the problem greatly.

Matter and curvature perturbations, that is $\delta T^0_0/\bar T^0_0$ and $\delta R/\bar R$, and particle velocities are kept at all orders, which allows us to study non-linear structure formation and relativistic or quasi-relativistic particle motion, such as CDM that has undergone rare extreme accelerations, and also intrinsically relativistic species like massive neutrinos.

Metric perturbations are assumed very small and are normally kept at first order. However, we also keep terms quadratic in $\Phi$ provided that they contain two space derivatives (e.g. $\Phi\Delta\Phi$, etc.), which makes sense since the Poisson equation (not exact in GR nor $f(R)$, but qualitatively accurate nonetheless) dictates $\partial^2\Phi \propto \delta\rho$ and we are keeping $\delta\rho/\bar\rho$ at all orders.

Our approximation scheme is summarised in Table~\ref{tab_orders_GR_fR}. We assume as usual that $\bar f_R \ll 1$ as well as $\delta f_R\ll 1$, although the latter can be of the same order of magnitude and even larger than $\bar f_R$, and could be a ``large'' perturbation in the same sense as $\Phi$. The scalaron appears in the equations in a way similar to that of $\Phi$, contributing to the fifth-force effects of $f(R)$, and also directly sources the gravitational slip $\chi$ (see~\S\ref{sec_chi_and_xi}), so we might risk losing potentially important features of the solutions by neglecting terms containing it.

Furthermore, we follow the standard approach and work in the quasi-static approximation, which consists in neglecting time derivatives. Therefore, we neglect $\delta \ddot f_R $, but unlike in other works which are based on the strictly Newtonian version of $f(R)$, we do keep the term proportional to $\H\,\delta \dot f_R $, as we do for $\Phi$ (see~\S\ref{sec_00_phi}).

\begin{table}[t]
	\centering
	\begin{tabular}{c c}
		Quantity & Order\\
		\hline
		\vspace*{-1.25em}
		\\
		$\Phi, \dot\Phi, \ddot\Phi, \delta f_R, \delta \dot f_R, \chi, \dot\chi, \ddot\chi$                 & $\epsilon$ \\
		$\Phi_{,i}, \dot\Phi_{,i}, \delta f_{R,i}, \delta \dot f_{R,i}, \chi_{,i}, \dot\chi_{,i}$                        & $\sqrt\epsilon$ \\
		$\Phi_{,ij}, \delta f_{R,ij}, \chi_{,ij}$                                                        & 1 \\
		$B_i$, $\dot B_i$, $\ddot B_i$, $B_{i,j}$, $\dot B_{i,j}$, $\ddot B_{i,j}$        & $\epsilon$ \\
		$h_{ij}$, $\dot h_{ij}$, $\ddot h_{ij}$, $h_{ij,k}$, $\dot h_{ij,k}$, $h_{ij,k\ell}$ & $\epsilon$ \\
		$\delta T^0_0/\bar T^0_0$, $\delta R/\bar R$                                       & 1 \\
		$T^0_i/\bar T^0_0$                                                                & $\sqrt\epsilon$ \\
		$\Pi_{ij}/\bar T^0_0$                                                             & $\epsilon$ \\
		$v^i$, $q_i$                                                                      & 1 \vspace*{.25em}\\
		\hline

	\end{tabular}
	\caption{Orders of approximation used. The various fields are defined in~\S\ref{sec_basics}.}
	\label{tab_orders_GR_fR}
\end{table}

\subsection{Scalaron Field: Trace Equation}
\label{sec_trace_equation}

We begin by considering the trace equation to update the scalaron field. Following the prescriptions in Tab.\ref{tab_orders_GR_fR}, we obtain
\begin{equation}
	\label{eq_pert_trace}
	(1+2\Phi)\Delta\delta f_R  - 2 \H\delta\dot f_R  + \frac{a^2}{3} \left[ f_R \delta R - 2\delta f + \bar R \delta f_R \right] = \frac{a^2}{3}(\delta R + 8\pi G \delta T)\,,
\end{equation}
Keep in mind that $\delta R = \delta R(f_R)$ and $\delta f = \delta f(f_R)$. This equation is a convenient choice as the first equation to solve in the code (after updating the energy-momentum tensor) because metric perturbations (and in fact only $\Phi$) enter the equation only in a sub-leading term ($\Phi \Delta\delta f_R  \ll \Delta\delta f_R$), so the error we make in using the old value of $\Phi$ will be negligible.

Note that the term containing $\delta \dot f_R $ is dealt with numerically by splitting it as
\begin{equation}
	\label{eq_xi_dot}
	\delta \dot f_R = \delta \dot f_R^{t-\frac{1}{2}} = \frac{\delta f_R^t - \delta f_R^{t-1}}{\dd\tau}\,.
\end{equation}
The index $t$, though it has an obvious correspondence with the cosmological time, is simply a discrete index labelling the simulation steps.

\subsubsection{Relaxation Solver}
The difficulty in solving~\eqref{eq_pert_trace} comes from the fact that the relation between $\delta f_R$ and $\delta R$ is in general (highly) non-linear so we cannot use standard spectral methods (e.g. FFT), rather we have to rely on relaxation methods. Schematically, we start from an equation of the form
\begin{equation}
	\label{eq_schematic_qs}
	U[\delta f_R] = S(T_\mn,\delta f_R) \so Y[\delta f_R,T_\mn] \equiv U - S = 0\,.
\end{equation}
In~\eqref{eq_schematic_qs}, $U$ is a non-linear differential operator acting on $\delta f_R$, and the source term $S$ can in principle contain terms depending on $\delta f_R$ as well. Starting from an initial guess $\delta f_R = \delta f_R^{(0)}$, associated with a residual $r^{(0)}$
\begin{equation}
	r^{(0)} \equiv Y[\delta f_R^{(0)}]\,,
\end{equation}
we implement a \textit{Newton-Raphson} iterative method, defined by
\begin{equation}
	\label{eq_xi_next_guess}
	\delta f_R^{(n+1)} = \delta f_R^{(n)} + \varepsilon^{(n)} \equiv \delta f_R^{(n)} - \left.\frac{Y}{\partial Y/\partial (\delta f_R)}\right|_{\delta f_R = \delta f_R^{(n)}}\,.
\end{equation}
where the error $\varepsilon^{(n)}$ quantifies the correction between values of $\delta f_R$ at consecutive iterations. Convergence is reached comparing the residual of the equation with some (small) pre-determined constant $r_{\rm c}$:
\begin{equation}
	||Y^{(n)}|| < r_{\rm c}\,,
\end{equation}
where typically $||\cdot||$ denotes $L_2$ norm taken over the whole grid:
\begin{equation}
	||Y^{(n)}||^2 = \sum_{i,j,k} (Y^{(n)}_{i,j,k})^2\,.
\end{equation}
Notice that the index $n$ denotes a progression in the relaxation, not in cosmological time. To clarify, Eq.~\eqref{eq_xi_next_guess} defines a sequence which progressively approaches the solution of~\eqref{eq_schematic_qs} at each time step. In this sense, an additional (fixed) index $t$ is implied in each quantity in~\eqref{eq_xi_next_guess}.

There are many possible sweeping strategies, the most popular and one of the easier to parallelise is probably the \textit{red-black} scheme, in which one solves the equation for cells of the same colour as in the colours of a chess board (straightforwardly generalised to 3 spatial dimensions), and then solves for the remaining half. The reason why this scheme is particularly useful for parallelisation is that when discretised on a lattice labelled by the indices $i,j,k$ and having cell size $\ell$, Eq.~\eqref{eq_pert_trace} only depends on the local value of $\delta f_R$ and of its Laplacian, which is computed from the nearest neighbours:
\begin{equation}
	\Delta \delta f_R^{i,j,k} = \frac{\delta f_R^{i+1,j,k} + \delta f_R^{i-1,j,k} + \delta f_R^{i,j+1,k} + \delta f_R^{i,j-1,k} + \delta f_R^{i,j,k+1} + \delta f_R^{i,j,k-1} - \delta f_R^{i,j,k}}{6 \ell^2}.
\end{equation}
Because the nearest neighbours of a black cell are red and vice versa, we can parallelise the update of all cells of one colour and afterwards update the remaining half.

Despite parallelising the relaxation, convergence often becomes increasingly slower as one approaches the exact solution. Formally, the issue is that the modes in the residual that have wavelengths longer than the grid size decrease more slowly than those with wavelengths comparable with the grid size. Therefore, especially for large grids, one often relies on multi-grid methods to speed up the convergence. Multi-grid algorithms speed up the convergence of these long wavelength modes by solving the equation on coarser grids (larger grid size), whose small-scale modes (in units of the grid size) correspond to larger scale modes in the finer grids.

For simplicity, we briefly illustrate the algorithm for two grids, but it can be easily generalised (see e.g.~\cite{Press:2007:NRE:1403886} for additional details). After a number of relaxation steps on the finer grid $\ell$, which produce a guess $\delta f_R^\ell$ with residual
\begin{equation}
 r^\ell = Y^\ell[\delta f_R^\ell]\,,
\end{equation}
we move to the coarser grid having cell size $L$ (typically $L = 2\ell$) using the \textit{restriction} operator%
\footnote{%
The restriction/injection/fine-to-coarse and the prolongation/interpolation/coarse-to-fine operators define the mapping between fields on two grids with different cell sizes. We choose a tri-linear interpolation for the prolongation operator, and its adjoint or inverse (full-weighting) for the restriction. See e.g.~\cite{Press:2007:NRE:1403886} for details.
}
$\mathcal R_{\ell\to L}$ on the scalaron and on the residual:
\begin{equation}
		\delta f_{R,\rm old}^L = \mathcal R_{\ell\to L}(\delta f_{R,\rm old}^\ell)\,, \qquad
		r_{\rm old}^L = \mathcal R_{\ell\to L}(r_{\rm old}^\ell)\,.
\end{equation}
On the new grid, we perform additional relaxations steps solving the modified equation
\begin{equation}
	Y^L[\delta f_{R,\rm new}^L] = Y^L[\delta f_{R,\rm old}^L] - r_{\rm old}^L\,.
\end{equation}
We then \textit{prolong} the error $\varepsilon^L \equiv \delta f_{R,\rm new}^L - \delta f_{R,\rm old}^L$ from the coarser grid to the finer grid, and thus correct the guess for $\delta f_R$ on the latter:
\begin{equation}
	\delta f_{R,\rm new}^\ell = \delta f_{R,\rm old}^\ell + \mathcal P_{L\to\ell}(\varepsilon^L)\,.
\end{equation}
Additional relaxation steps are then performed on the finer grid, and if required the multigrid cycle can be repeated until the desired precision is achieved.

\subsubsection{Change of variable}
\label{sec_xi_u}

Depending on the specific $f(R)$ model, $\delta f_R$ might only have a finite range of ``healthy'' values%
\footnote{%
This is not a generic feature of $f(R)$ models, but it occurs in several models designed to produce cosmic acceleration, including the Hu-Sawicki model. Typically, this happens whenever $f(R)$ and/or the relation $R \leftrightarrow f_R$ are ill-defined for $R<0$, or when the unbounded interval $-\infty < R < \infty$ is mapped into a bounded interval for $f_R$.
}%
, and it may happen that the sequence~\eqref{eq_xi_next_guess} accidentally pushes $\delta f_R$ outside this range. For instance, in the Hu-Sawicki model~\cite{Hu:2007nk}, the relation $R(f_R)$ is well-defined only for a definite sign of $f_R$, so clearly only a finite range of values of $\delta f_R \equiv f_R - \bar f_R$ is allowed. Where needed, as suggested in~\cite{Oyaizu:2008sr}, we circumvent this problem by using the auxiliary variable $u$, defined as
\begin{equation}
	\label{eq_u_xi}
	f_R \equiv \bar f_R e^u \quad \Leftrightarrow \quad \delta f_R = \bar f_R(e^u - 1)\,.
\end{equation}
This is of course not the only possible choice, and in principle each model should be considered individually. Once we have a relation $R(u)$ that is well-defined on the whole real axis, we can convert $\delta f_R \to u$ on each lattice point, re-formulate the trace equation in terms of $u$, and apply a strategy analogous to~\eqref{eq_xi_next_guess}, before converting back $u\to \delta f_R$. For the specific case~\eqref{eq_u_xi}, we have for example
\begin{equation}
	\delta f_R^{(n)} + \varepsilon^{(n)} = \bar f_R \left(e^{u^{(n)} + \delta u^{(n)}} -1 \right)\,,
\end{equation}
so expanding the right-hand side at first order in $\delta u$ (we drop the subscript $(n)$ for simplicity) we obtain
\begin{equation}
	\begin{aligned}
		\delta f_R + r \simeq \bar f_R\left( e^{u} + e^{u} \delta u - 1 \right) &= \delta f_R + \bar f_R e^{u} \delta u \\
		& = \delta f_R + f_R \delta u
	\end{aligned}
	\so
	\delta u \simeq \frac{\varepsilon}{f_R}\,.
\end{equation}
In practice, we can skip converting between $\delta f_R$ and $u$ using the following sequence
\begin{equation}
	\delta f_R^{(n+1)} = \bar f_R\left[\exp\left( u^{(n+1)}\right) - 1\right] \simeq f_R^{(n)} \exp \left(\frac{\varepsilon^{(n)}}{f_R^{(n)}}\right) - \bar f_R\,,
\end{equation}
where for clarity we remind the reader that $\varepsilon^{(n)}$ is given by~\eqref{eq_xi_next_guess}.

\subsection{Gravitational Potential: \texorpdfstring{$00$}{00} Equation}
\label{sec_00_phi}

Having solved the trace equation, we use the 00 equation to solve for $\Phi$, putting all terms containing the scalaron in the source term:
\begin{equation}
	\label{eq_fR_pert_00}
	\begin{aligned}
		\left(\Delta - \frac{3\H}{\dd\tau} - 3\H^2\right)\Phi_t &= -4\pi G a^2 \left(1 - 4\Phi - f_R\right) \delta T^0_0 + \frac{1 - 2\Phi - f_R}{2} \Delta\delta f_R \, - \\
		& \quad - \frac{3}{2}\H^2 \left(2\chi + \delta f_R  \right) -\frac{1}{2} \delta^{ij}\Phi_{,i}(3\Phi + \delta f_R )_{,j} \, + \\
		& \quad + \frac{R \delta f_R  + \bar f_R\delta R - \delta f}{4}a^2 - \frac{3\H}{\dd\tau}\Phi - \frac{3\H}{2}\delta \dot f_R  \\
		& \equiv S^0_0\,,
	\end{aligned}
\end{equation}
where it is implied that $\Phi = \Phi_{t-1}$ and $\chi = \chi_{t-1}$ in the right-hand side. With this new source term, $\Phi_t$ is readily computed via
\begin{equation}
	\tilde\Phi_t = -\left(k^2 + \frac{3\H}{\dd\tau} + 3\H^2\right)^{-1} \tilde S_0^0\,,
\end{equation}
which allows us to update both $\Phi_t$ and $\dot\Phi_{t-\frac{1}{2}}$ analogously to~\eqref{eq_xi_dot}.

\subsection{Vector Field (Elliptic Constraint): \texorpdfstring{$0i$}{0i} Equation}
\label{sec_elliptic}

The $0i$ equation
\begin{equation}
	\label{eq_fR_pert_0i}
	\begin{aligned}
		& -\frac{1}{2}\Delta B_i - B_i \Delta\Phi + \delta^{jk}B_{j}(\delta f_R  - \Phi)_{,ik} - \H(2\Phi - 2\chi + \delta f_R)_{,i}\, - \\
		&\hspace{10em}  - \, 2\dot\Phi_{,i} + \delta \dot f_{R,i} - \dot f_R\Phi_{,i} -2 (\Phi - \chi) \delta \dot f_{R,i} = 8\pi G a^2 T^0_i\,,
	\end{aligned}
\end{equation}
can be used to evolve the vector mode $B_i$ via an elliptic constraint equation. Projecting~\eqref{eq_fR_pert_0i} on the spin-1 component, using the operator
\begin{equation}
	\label{eq_projector_spin_1}
	P_{(1)}^{ij} \equiv k^2\delta^{ij} -k^i k^j\,,
\end{equation}
we obtain
\begin{equation}
	\label{eq_update_B_0i}
		\delta^{ij}(\tilde{B_i})_t = 2\,k^{-4}\left(k^2\delta^{ij} - k^i k^j\right)\,{\rm Fourier}\left\{8\pi G a^2 T^0_i + B_i\Delta\Phi + \delta^{k\ell} B_{k}(\Phi - \delta f_R)_{,i\ell}\right\}\,.
\end{equation}

\subsection{Traceless \texorpdfstring{$ij$}{ij} Equation}
We finally consider the traceless part of the $ij$ equations, namely
\begin{equation}
	\label{eq_fR_pert_ij}
	\begin{aligned}
		& \left(\delta^i_k\delta^j_\ell - \frac{1}{3}\delta_{k\ell}\delta^{ij}\right) \left[\dot B_{(i,j)} + 2\H B_{(i,j)} + 2\Phi_{(,i}(\Phi - \delta f_R)_{,j)} \right. + \\
		&\hspace{.2\textwidth} +  \left. 2(2\Phi - \chi)\Phi_{,ij} - (1+2\Phi)\delta f_{R,ij} + \chi_{,ij} \right] = 8\pi G a^2 \Pi_{k\ell}\,,
	\end{aligned}
\end{equation}
where
\begin{equation}
	\Pi_{ij} \equiv \left( \delta_{ik}\delta^\ell_j - \frac{1}{3}\delta^\ell_k\delta_{ij} \right) T^k_\ell\,,
\end{equation}
and round brackets in indices denote symmetrisation:
\begin{equation}
	A_{(i}B_{j)} \equiv \frac{A_iB_j + A_jB_i}{2}\,.
\end{equation}
This equation will be used to evolve the gravitational slip $\chi$ (via its spin-0 projection), and possibly $B_i$ (via the spin-1 projection) through a parabolic equation.

As mentioned previously, we are neglecting the tensor perturbations $h_{ij}$. They would enter these equations through a term proportional to $\ddot h_{ij} + 2\H \dot h_{ij} - \Delta h_{ij}$, unchanged from GR to $f(R)$, inside the square brackets.

We move all non-linear terms and terms containing $\Phi$ and $\delta f_R$ (already updated at this point of the cycle) to the right-hand side and project on the traceless part, obtaining
\begin{equation}
	\label{eq_traceless_source}
	\begin{aligned}
		& \dot B_{(i,j)} + 2\H B_{(i,j)} + \chi_{,ij} - \frac{1}{3}\delta_{ij}\Delta\chi = \\
		& \qquad\qquad = 8\pi G a^2 \Pi_{ij} - \left(\delta^k_i\delta^\ell_j - \frac{1}{3}\delta^{k\ell}\delta_{ij}\right)\left[2\Phi_{,k}\Phi_{,\ell} + 2(2\Phi - \chi)\Phi_{,k\ell} - \delta f_{R,k\ell}\right] \\
		& \qquad\qquad \equiv S_{ij} - \frac{1}{3}\delta_{ij} S + \delta f_{R,ij} - \frac{1}{3}\delta_{ij}\Delta\delta f_R\,.
	\end{aligned}
\end{equation}
We kept $\delta f_R$ separated from the rest of the source term in the right-hand side for reasons that will be clear shortly. Note also that
\begin{equation}
	\dot {\mathcal B_i} \equiv \partial_\tau(a^2 B_i) = a^2(\dot B_i + 2\H B_i)\,,
\end{equation}
so this is going to be the combination that is actually used to solve this equation. In Fourier space, we obtain
\begin{equation}
	ia^{-2}\dot{\tilde{\mathcal B}}_{(i}k_{j)} - k_i k_j\tilde\chi + \frac{k^2}{3}\delta_{ij} \tilde\chi = \tilde S_{ij} - \frac{1}{3}\delta_{ij}\tilde S - k_i k_j \delta\tilde f_R + \frac{k^2}{3}\delta_{ij}\delta \tilde f_R \,.
\end{equation}

\subsubsection{Spin-0 Mode}
\label{sec_chi_and_xi}

We first calculate $\chi$ by projecting on the spin-0 part, using the projection operator
\begin{equation}
	P_{(0)}^{ij} \equiv k^2\delta^{ij} - 3k^i k^j\,,
\end{equation}
which yields
\begin{equation}
	2k^4\tilde\chi^t = \left(k^2 \delta^{ij} - 3k^i k^j\right)\tilde S_{ij}\left( \Phi^t, \chi^{t-1} \right) + 2k^4\delta \tilde f_R^t \,,
\end{equation}
so we finally obtain
\begin{equation}
	\label{eq_update_chi_final}
	\chi^t = \delta f_R^t + {\rm Fourier}^{-1}\left\{ \frac{1}{2k^4}\left(k^2 \delta^{ij} - 3k^i k^j\right)\tilde S_{ij}\left( \Phi^t, \chi^{t-1} \right) \right\}\,.
\end{equation}
Notably, we can avoid the computation of the scalaron-dependent terms in the source, and simply add $\delta f_R$ to the final result, which is why we chose to keep those terms explicit in~\eqref{eq_traceless_source}.

Eq.~\eqref{eq_update_chi_final} is essentially showing how the new scalar is a source of anisotropic stress in $f(R)$ gravity theories. In fact, if one assumes that $\Phi$ and $T_\mn$ are essentially the same as in GR\footnote{This is obviously not a good quantitative approximation, but it helps in understanding the qualititative effect of modified gravity on $\chi$.}, then it is the \textit{difference} $(\chi - \delta f_R)$ that is roughly equal to $\chi_{\rm GR}$ (see also~\S\ref{sec_newtonian_limit}), or equivalently
\begin{equation}
	\chi_{f(R)} \approx \chi_{\rm GR} + \delta f_R\,.
\end{equation}

\subsubsection{Spin-1 Mode}
\label{sec_parabolic}

Projecting on the spin-1 component, using the projector~\eqref{eq_projector_spin_1} through the contraction
\begin{equation}
	P_{(1)}^{i\ell}k^j\left(\rm Equation\right)_{ij}\,,
\end{equation}
where $P_{(1)}$ was defined in~\eqref{eq_projector_spin_1}, finally yields, using the gauge condition $\tilde{\mathcal B_i} k^i = 0$,
\begin{equation}
	\dot{\tilde{\mathcal B}_i} = -\frac{2ia^2}{k^4}\delta_{i\ell}\left(k^2\delta^{j\ell} - k^j k^\ell\right)k^m\tilde S_{jm}\,.
\end{equation}
The field $\tilde{\mathcal B}_i$ is then updated with a simple Euler criterion
\begin{equation}
	\label{eq_update_B_ij_final}
	\tilde{\mathcal B}_i^t = \tilde{\mathcal B}_i^{t-1} - \frac{2ia^2\dd\tau}{k^4}\delta_{i\ell}\left(k^2\delta^{j\ell} - k^j k^\ell\right)k^m\tilde S_{jm}\,.
\end{equation}

\section{The Newtonian Limit}
\label{sec_newtonian_limit}

The Newtonian limit of~\eqref{eq_fR_pert_00} is essentially the $f(R)$ equivalent of the Poisson equation, which is
\begin{equation}
	\Delta\Phi_N = 4\pi G a^2 \delta\rho\,,
\end{equation}
where $\Phi_N$ is the Newonian potential. Furthermore, the geodesic motion of non-relativistic test particles can be approximated by
\begin{equation}
	\ddot{\mathbf x} = \ddot{\mathbf x}_{\rm GR} \equiv -\nabla \Phi_{\rm GR}\,.
\end{equation}
Several fundamental assumptions are being made in the Newtonian limit, namely that $\Phi_N$ is small so that the first order terms suffice, and that we are in the deep sub-horizon, quasi-static regime so that
\begin{equation}
	\partial_t, \H \ll k\,.
\end{equation}
We should also assume that $\delta P \ll \delta\rho$, as is the case if velocities are non-relativistic. Moreover, we are assuming that only the leading corrections to GR are relevant, which allows us to get rid of plenty of terms such as $f_R\ll 1$, $f_{RR} R \ll 1$ and so on. With these approximations, we find that the leading contribution to~\eqref{eq_fR_pert_00} is given by
\begin{equation}
	\label{eq_fR_00_Newtonian}
	\Delta\Phi = 4\pi G a^2\delta\rho + \frac{1}{2}\Delta\delta f_R\,,
\end{equation}
so that the correction to the field $\Phi$, assuming that $\delta\rho$ evolves practically as in GR, is roughly
\begin{equation}
	\Phi_{f(R)} \approx \Phi_{\rm GR} + \frac{1}{2}\delta f_R\,.
\end{equation}
Moreover, as we have seen in~\S\ref{sec_chi_and_xi}, the gravitational slip $\chi$ (which vanishes identically in Newtonian gravity) is now sourced directly by the scalaron, so that
\begin{equation}
	\chi \approx \chi_{\rm GR} + \delta f_R\,.
\end{equation}
These equations provide us with one the more intuitive ways to see how the additional scalar sources the gravitational potential and contributes as a fifth force, in fact the acceleration of a test particle will be
\begin{equation}
	\ddot{\mathbf x} \simeq - \nabla \Psi = -\nabla(\Phi - \chi) \approx - \nabla \Phi_{\rm GR} + \frac{1}{2}\nabla \delta f_R = \ddot{\mathbf x}_{\rm GR} + \delta\ddot{\mathbf x}_{f(R)} \,.
\end{equation}
Similarly, the Newtonian limit of the trace equation~\eqref{eq_pert_trace} reads
\begin{equation}
	\label{eq_fR_trace_Newtonian}
	\Delta\delta f_R = \frac{a^2}{3}(\delta R - 8\pi G \delta \rho),
\end{equation}
which in combination with the previous results yields
\begin{equation}
	\Delta\Phi = a^2\left(\frac{8\pi G}{3}\delta\rho + \frac{\delta R}{6}\right), \qquad \Delta\Psi = a^2\left(\frac{16\pi G}{3}\delta\rho - \frac{\delta R}{6}\right),
\end{equation}
which are precisely the equations used in the pioneering~\cite{Oyaizu:2008sr}.

In the Newtonian approximation, we replace~\eqref{eq_pert_trace} and~\eqref{eq_fR_pert_00} with~\eqref{eq_fR_trace_Newtonian} and~\eqref{eq_fR_00_Newtonian}, respectively. When computing the particle dynamics, we moreover neglect $B_i$ and $\chi$, as well as the standard relativistic corrections (typically of order $v^2/c^2 \ll 1$). See the original~\texttt{gevolution} paper~\cite{Adamek:2016zes} for details.

\section{Results}

In this section we will present some results from our simulations and their comparison with $\Lambda$CDM and existing modified gravity codes~\cite{Li:2012by, Puchwein:2013lza}. A more detailed discussion will appear in a following paper~\cite{Reverberi_Davero_et_al_preparation}.

\subsection{Point Mass}

\begin{figure}[t!]
  \centering
  \begin{overpic}[width=.5\textwidth]{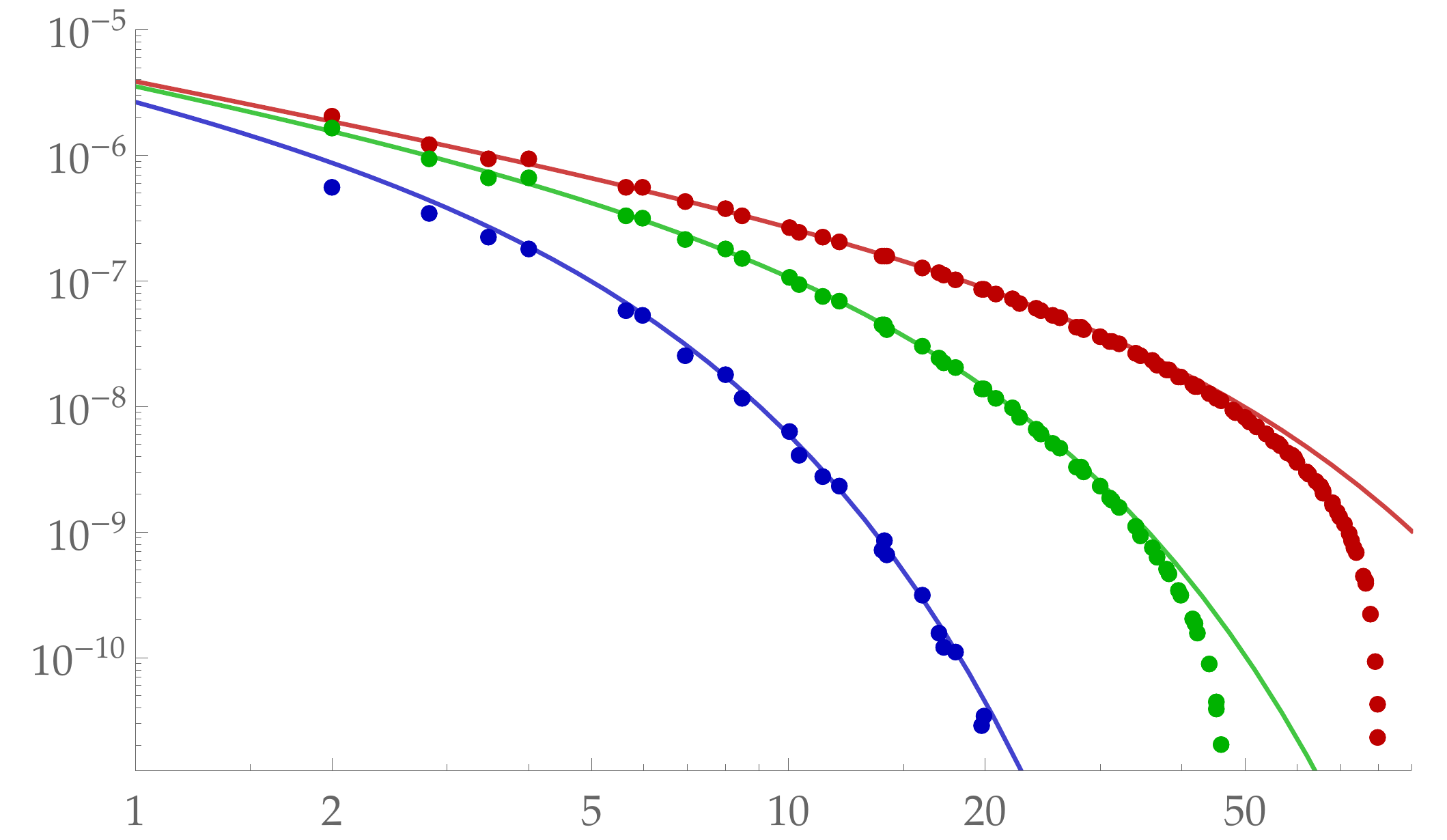}
    \fontsize{8}{0}\selectfont
    \put(46,-3.5)
    {
    $r$ [Mpc/$h$]
    }
    \put(-12,30)
    {
    $\delta f_R(r)$
    }
    \put(12,62)
    {
    $
    \displaystyle
    {\color{blue} |f_{R_0}| = 10^{-6}}
    \qquad
    {\color{green!50!black}  |f_{R_0}| = 10^{-5}}
    \qquad
    {\color{red} |f_{R_0}| = 10^{-4}}
    $
    }
  \end{overpic}
  \vspace*{1em}
  \caption{Point-mass solutions and comparison with the analytical prediction~\eqref{eq_BH_analytical}. Deviations at large radii are due to our periodic conditions whereas~\eqref{eq_BH_analytical} assumes asymptotically flat boundary conditions.}
  \label{fig_BH_HS}
\end{figure}

As a first test, we consider the static field produced by a point mass located in the centre of a (256Mpc/$h$)$^3$ cubic box (with periodic boundary conditions), solving on $128^3$ grid points (hence the spatial resolution is $\ell_{\rm cell} = 2$Mpc/$h$), and compare these results with the analytical solutions obtained linearising~\eqref{eq_fR_trace_Newtonian}:
\begin{equation}
  \label{eq_trace_linearised}
  \Delta \delta f_R = \frac{\delta f_R}{3\bar f_{RR}} - \frac{8\pi G}{3} \delta\rho \,,
\end{equation}
with a density field
\begin{equation}
  \label{eq_BH_density}
  \delta\rho =
  \begin{cases}
    10^{-4}(N^3 - 1)\bar\rho & \text{point mass cell} \\
    - 10^{-4} \bar\rho & \text{elsewhere}
  \end{cases}
\end{equation}
The formal solution for an actual point mass $\rho = m\,\delta^{(3)}(\mathbf r)$ in an asymptotically flat Universe is trivially a Yukawa-like profile
\begin{equation}
  \label{eq_BH_analytical}
  \delta f_R = \frac{2 Gm}{3}\frac{e^{-r/\sigma}}{r}\,,
\end{equation}
where
\begin{equation}
  \sigma^2 = 3 \bar f_{RR} \,.
\end{equation}
The conversion between $m$ and~\eqref{eq_BH_density} is given by
\begin{equation}
  m \to 10^{-4}(N^3 - 1)\bar\rho V_{\rm cell}\,,
\end{equation}
where $V_{\rm cell} = \ell_{\rm cell}^3$ is the volume of a lattice cell. For $|f_{R_0}| = 10^{-6}$, we replace the point mass with a Gaussian profile
\begin{equation}
  \rho \propto \exp \left(-\frac{r^2}{\ell^2_{\rm cell}}\right)\,,
\end{equation}
because $\delta f_R > |f_{R_0}|$ at small distances, and hence the linear approximation fails. In all cases, we expect solutions to deviate from the analytical result nearest to the overdensity and to the boundaries of the box, due to finite resolution and finite size effects, respectively. Results are shown in Fig.~\ref{fig_BH_HS}.

\subsection{Cosmological Simulations}

We performed three simulations of a (512 Mpc$/h)^3$ comoving box with $512^3$ grid points and the same number of CDM particles, for different values of $|f_{R_0}| = 10^{-4},10^{-5},10^{-6}$. This allows us to study the limit in which the model reduces essentially to $\Lambda$CDM and when instead deviations become significant. The other cosmological parameters used in the simulations are~\cite{Aghanim:2018eyx}: $h=0.6736$, $\Omega_b h^2 = 0.02237$, $\Omega_c h^2 = 0.1200$, $n_s = 0.9649$, $A_s = 2.099\times 10^{-9}$ (at 0.05 Mpc$^{-1}$). We should stress that these cosmological parameters are not the most up-to-date values available (see e.g.~\cite{Aghanim:2018eyx}) but were chosen for a direct comparison with the existing codes~\cite{Li:2012by, Puchwein:2013lza}. We plan on using more recent values in an upcoming publication~\cite{Reverberi_Davero_et_al_preparation} in which we discuss our results in more detail.

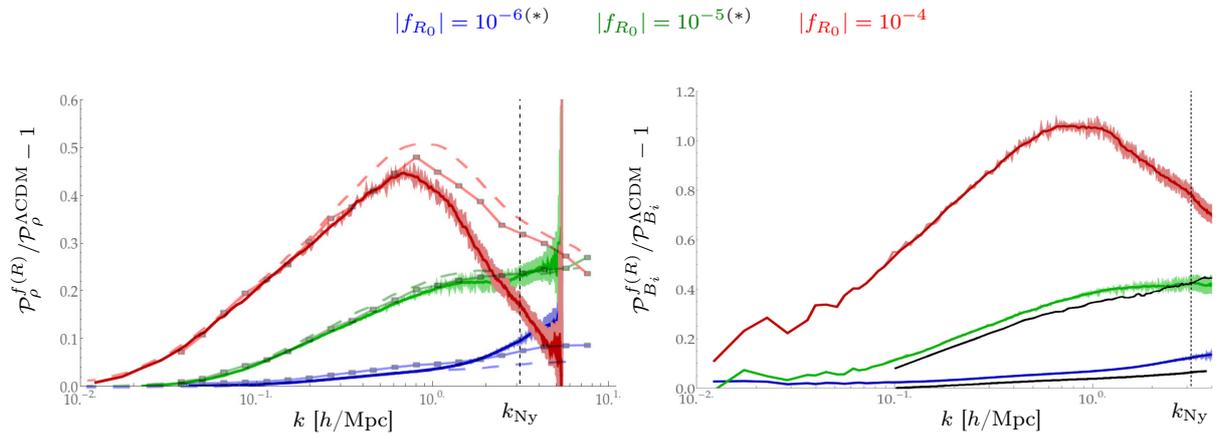
\begin{figure}[t]
  \centering
  \begin{subfigure}[t]{.48\textwidth}
    \hspace*{.5em}
    \begin{overpic}[width=1.\textwidth]{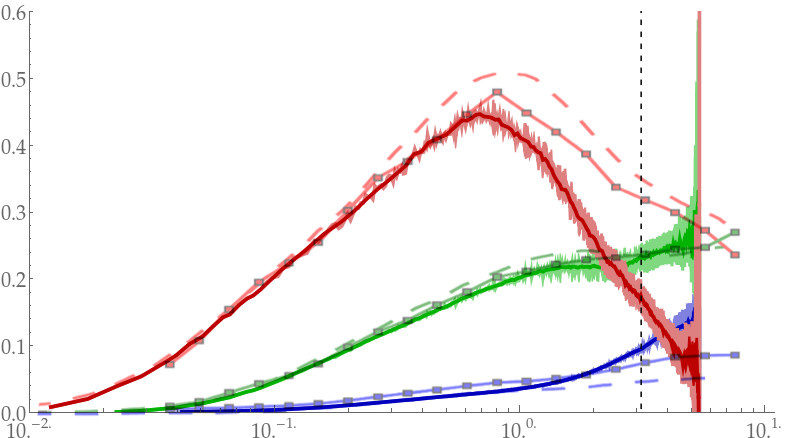}
      \fontsize{8}{0}\selectfont
      \put(40,-4)
      {
      $k$ [$h$/Mpc]
      }
      \put(-10,16)
      {
      \begin{sideways}
        $\mathcal P_{\rho}^{f(R)}/\mathcal P_{\rho}^{\Lambda \text{CDM} } - 1$
      \end{sideways}
      }
      \put(58,67)
      {
      $
      \displaystyle
      {\color{blue} |f_{R_0}| = 10^{-6}}^{(*)}
      \qquad
      {\color{green!50!black}  |f_{R_0}| = 10^{-5}}^{(*)}
      \qquad
      {\color{red} |f_{R_0}| = 10^{-4}}
      $
      }
      \put(77,-2.5)
      {
      $ k_{\rm Ny}$
      }
    \end{overpic}
    \vspace*{.1em}
    \caption{Results for the relative enhancement of the matter power spectrum compared to $\Lambda$CDM, compared to the results of~\cite{Li:2012by} (dashed lines) and~\cite{Puchwein:2013lza} (dots). The discrepancy for $|f_{R_0}| = 10^{-4}$ near the Nyquist scale $k_{\rm Ny}$ (dashed vertical line) most likely have no physical origin and are instead artifacts of unavoidable smoothing and finite resolution effects. Overall, our results and~\cite{Li:2012by,Puchwein:2013lza} agree extremely well for $k \lesssim 0.25\,k_{\rm Ny}$ even when small scale differences are largest.}
    \label{fig_matter}
  \end{subfigure}
  \hspace*{1.em}
  \begin{subfigure}[t]{.48\textwidth}
    \centering
    \hspace*{.5em}
    \begin{overpic}[width=.95\textwidth,height=4.2cm]{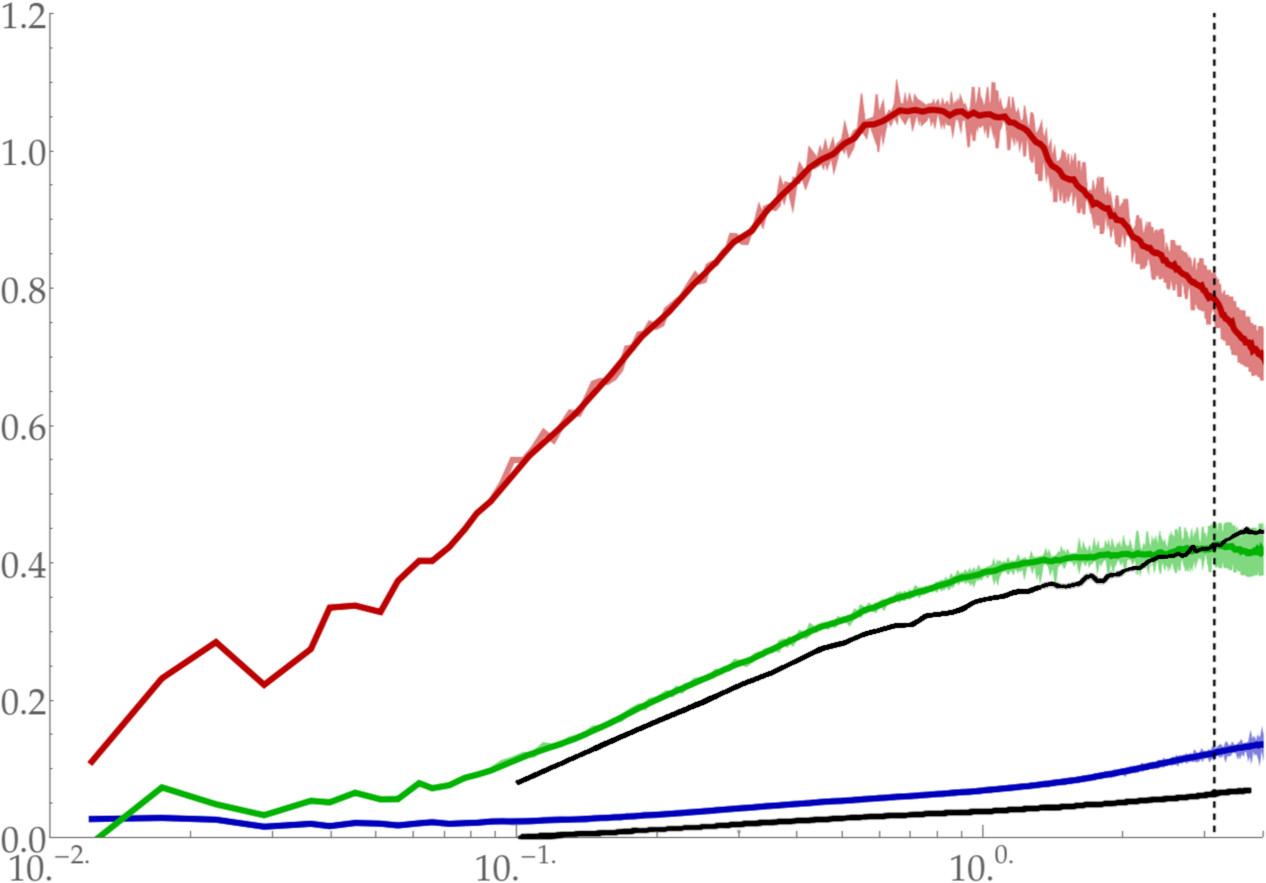}
      \fontsize{8}{0}\selectfont
      \put(48,-4)
      {
      $k$ [$h$/Mpc]
      }
      \put(-11,17)
      {
      \begin{sideways}
        $\mathcal P^{f(R)}_{B_i}/\mathcal P_{B_i}^{\Lambda \text{CDM}} - 1$
      \end{sideways}
      }
      \put(91,-2.5)
      {
      $ k_{\rm Ny}$
      }
    \end{overpic}
    \vspace*{.1em}
    \caption{Power excess for the vector modes $B_i$. Deviations are generically larger than those of $\delta \rho$ and follow a similar qualitative behaviour. As was remarked in the text, any departure from $\Lambda$CDM is not due to additional sources in the evolution equation for the vector modes, but to the indirect effect of changes in the scalar potentials and in the matter source. $^{(*)}\,$For a direct comparison with the results of~\cite{Thomas:2015dfa} (black solid lines), we used a slightly different cosmology than the other simulations and $|f_{R_0}| = 1.289\times 10^{-5}, 1.289\times 10^{-6}$.}
    \label{fig_vector}
  \end{subfigure}
  \caption{Results for $\delta\rho$ and $B_i$.}
\end{figure}

\subsubsection{Matter Power Spectrum}

We present out results for the matter power spectrum in Fig.~\ref{fig_matter}, where we overlay our power spectra to those of MG-gadget~\cite{Puchwein:2013lza}. We observe excellent agreement with~\cite{Puchwein:2013lza} at all scales except for $|f_{R_0}| = 10^{-4}$ (where deviations from GR are most significant), where our solutions have a faster drop in the power excess around $k \gtrsim 1 h/$Mpc. This is most likely due to the proximity to the Nyquist scale $k_{\rm Ny}$, and in general we do not expect our results to be accurate and competitive with those of MG-gadget so close to $k_{\rm Ny}$\footnote{Notice that the results of the original \texttt{gevolution} code also deviate from those of Gadget at small scales $k \simeq k_{\rm Ny}$, and that in particular \textit{less} power is produced around those scales (see~\cite{Adamek:2016zes} for details).}.

Altogether, we can still state that our results for $|f_{R_0}| = 10^{-4}$ agree with those of existing codes very well, even in the non-linear regime, at scales larger than about a factor of 4-5 times the Nyquist scale; the agreement is even better and essentially perfect at any scale $k < k_{\rm Ny}$ for $|f_{R_0}| = 10^{-5},10^{-6}$. Considering the intrinsic limitations of a fixed-grid approach compared to an adaptive mesh, we can consider this agreement very satisfactory.

\subsubsection{Vector Modes}

In figure~\ref{fig_vector} we present our results for the power spectrum of vector modes $B_i$. We also compare our results with those of~\cite{Thomas:2015dfa}, in which vector modes are computed in a Post-Friedmannian framework. For this reason, we used slightly different cosmological parameters (see~\cite{Thomas:2015dfa}) and the values $|f_{R_0}| = 10^{-4}$, $1.289\times 10^{-5}$ and $1.289\times 10^{-6}$.

We can see that the power excess compared to $\Lambda$CDM is larger by roughly 100\%, 30\% and a few percent for $|f_{R_0}|=10^{-4}$, $10^{-5}$ and $10^{-6}$ respectively, at $k\simeq 1 h/$Mpc. We should stress again that this excess is not due to any additional term appearing directly in the evolution equation for $B_i$, but indirectly due to how the scalar potentials (and its gradient, which sources the vector modes) and the matter source are modified because of the $f(R)$ contributions. It is therefore remarkable that the power excess is even larger than for $\Phi$ which is sourced by $\delta f_R$ directly. The agreement with~\cite{Thomas:2015dfa} is overall rather good but we do detect a slight excess of extra power. We plan to investigate this point further in a following publication.

\subsubsection{Curvature}

Next we present the solutions for the scalar curvature perturbations $\delta R$, in figure~\ref{fig_deltaR}. The figure shows the ratio of power spectra instead of the relative power excess as in the previous cases, because the differences from the $\Lambda$CDM solution
\begin{equation}
  \delta R = -8\pi G \delta T
\end{equation}
can be of several orders of magnitude and not at most of order unity as in the cases of matter and vector perturbations.

We see that the $\Lambda$CDM limit appears to be recovered in the appropriate limit as $|f_{R_0}|$ decreases, but we also notice that deviations are significant, especially at smaller scales, already at relatively small values of $|f_{R_0}|$ and rather strikingly for $|f_{R_0}| = 10^{-4}$ at basically all scales. While deviations in $\delta R$ are not easily testable alone, as the main cosmological observable is the matter power spectrum, these results suggest that even in those cases in which deviations in $\delta\rho$ and $\Phi$ are relatively small, the linear expansion around some reference curvature\footnote{Note that $\tilde R$ needs not be the cosmological background curvature $\bar R$, but can be any ``sensible'' choice, for instance the GR solution $\tilde R = - 8\pi G T$.} $\tilde R$ might give extremely inaccurate approximations to the real solution. For example, we can not assume
\begin{equation}
  f_R(\tilde R + \delta R) \simeq \tilde f_R + \tilde f_{RR} \delta R
\end{equation}
if $|\delta R/\tilde R| \sim 1$, and similarly for other derivatives. Moreover, this is typically exacerbated by the high non-linearity of the relation between $f_R$ and $R$ in $f(R)$ models relevant for cosmic acceleration

The conclusion is that one should be very careful when producing estimates for the effects of modified gravity in the approximation $R \simeq R^{\Lambda\text{CDM}}$, because this could be violated by many orders of magnitude even in those cases for which the gravitational potential and the matter power spectrum are not too different that in GR.

\subsubsection{Gravitational Slip}

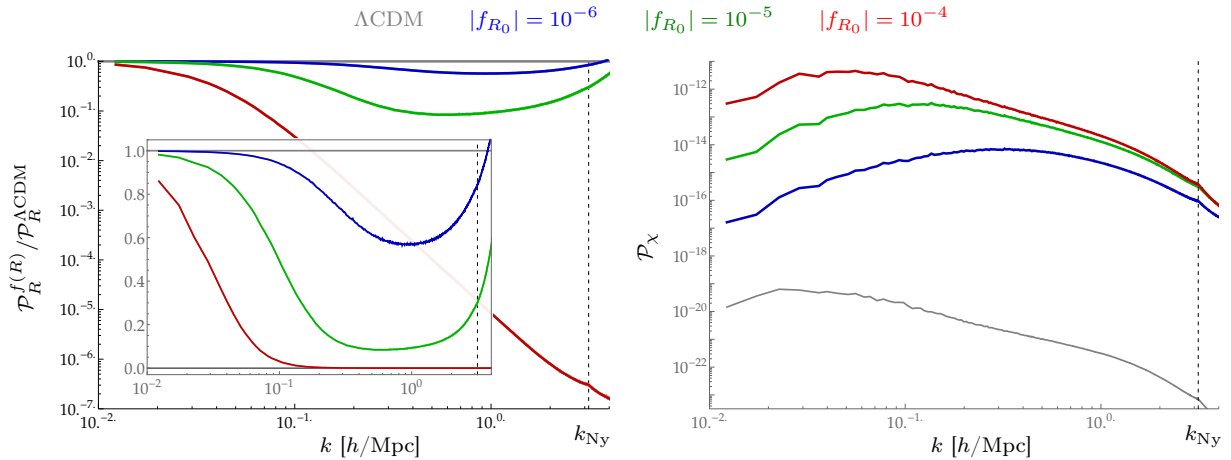
\begin{figure}[t]
  \begin{subfigure}[t]{.48\textwidth}
    \centering
    \begin{overpic}[width=1\textwidth]{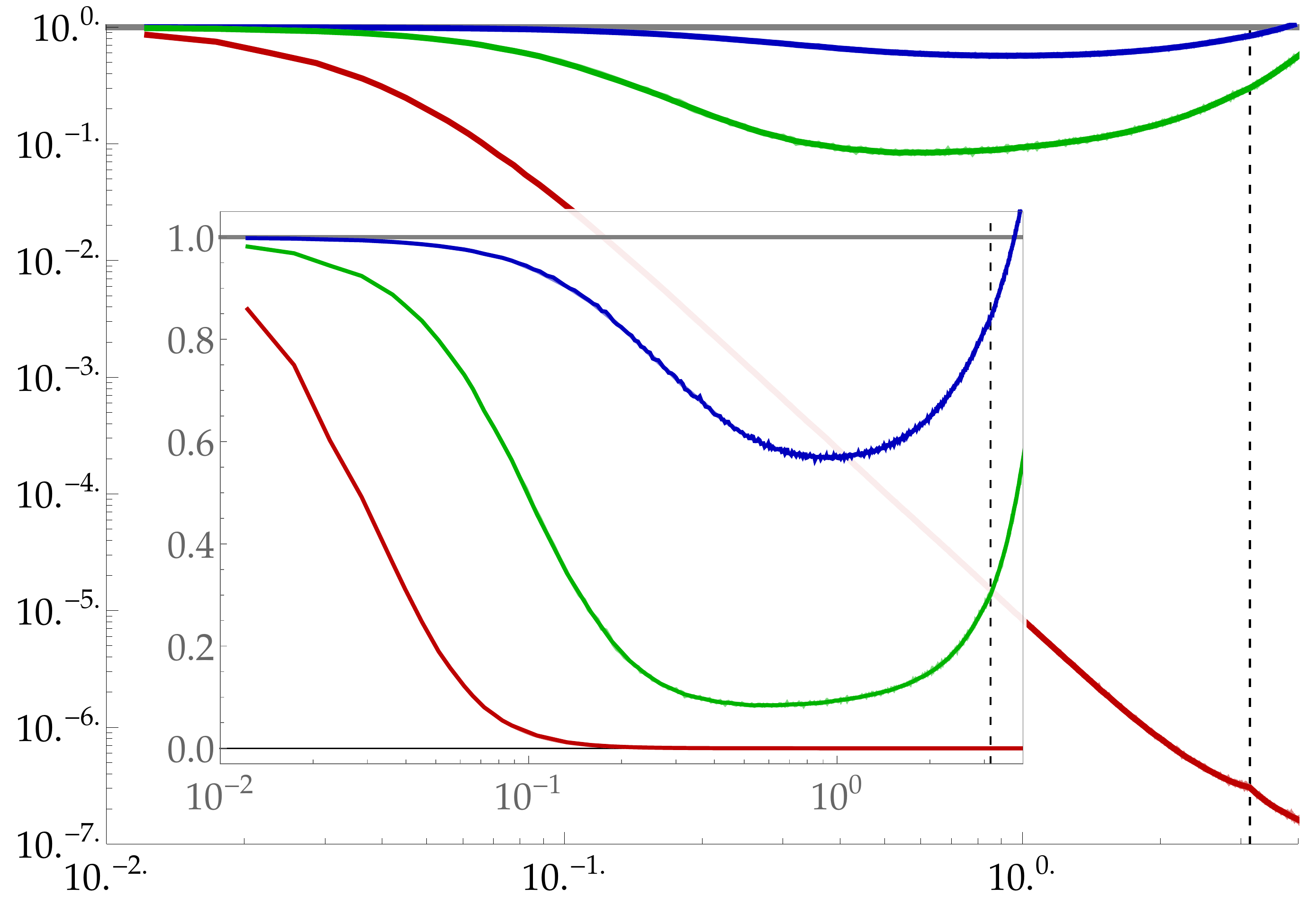}
      \fontsize{8}{0}\selectfont
      \put(46,-3)
      {
      $k$ [$h$/Mpc]
      }
      \put(-9,22)
      {
      \begin{sideways}
        $\mathcal P^{f(R)}_R/\mathcal P^{\Lambda\text{CDM}}_R$
      \end{sideways}
      }
      \put(52,73)
      {
      $
      \displaystyle
      {\color{gray} \Lambda\text{CDM}}
      \qquad
      {\color{blue} |f_{R_0}| = 10^{-6}}
      \qquad
      {\color{green!50!black}  |f_{R_0}| = 10^{-5}}
      \qquad
      {\color{red} |f_{R_0}| = 10^{-4}}
      $
      }
      \put(91,-1)
      {
      $ k_{\rm Ny}$
      }
    \end{overpic}
    \vspace*{.1em}
    \caption{Power spectra for the scalar curvature and comparison with the GR solution $\delta R = -8\pi G\delta T$. Although this solution is recovered in the appropriate limit, deviations are very significant even at relatively low $|f_{R_0}|$ (see the text for a discussion on this point). The two plots show the same results in logarithmic and linear scale on the vertical axis.}
    \label{fig_deltaR}
  \end{subfigure}
  \hspace*{1.em}
  \begin{subfigure}[t]{.48\textwidth}
    \centering
    \begin{overpic}[width=1\textwidth]{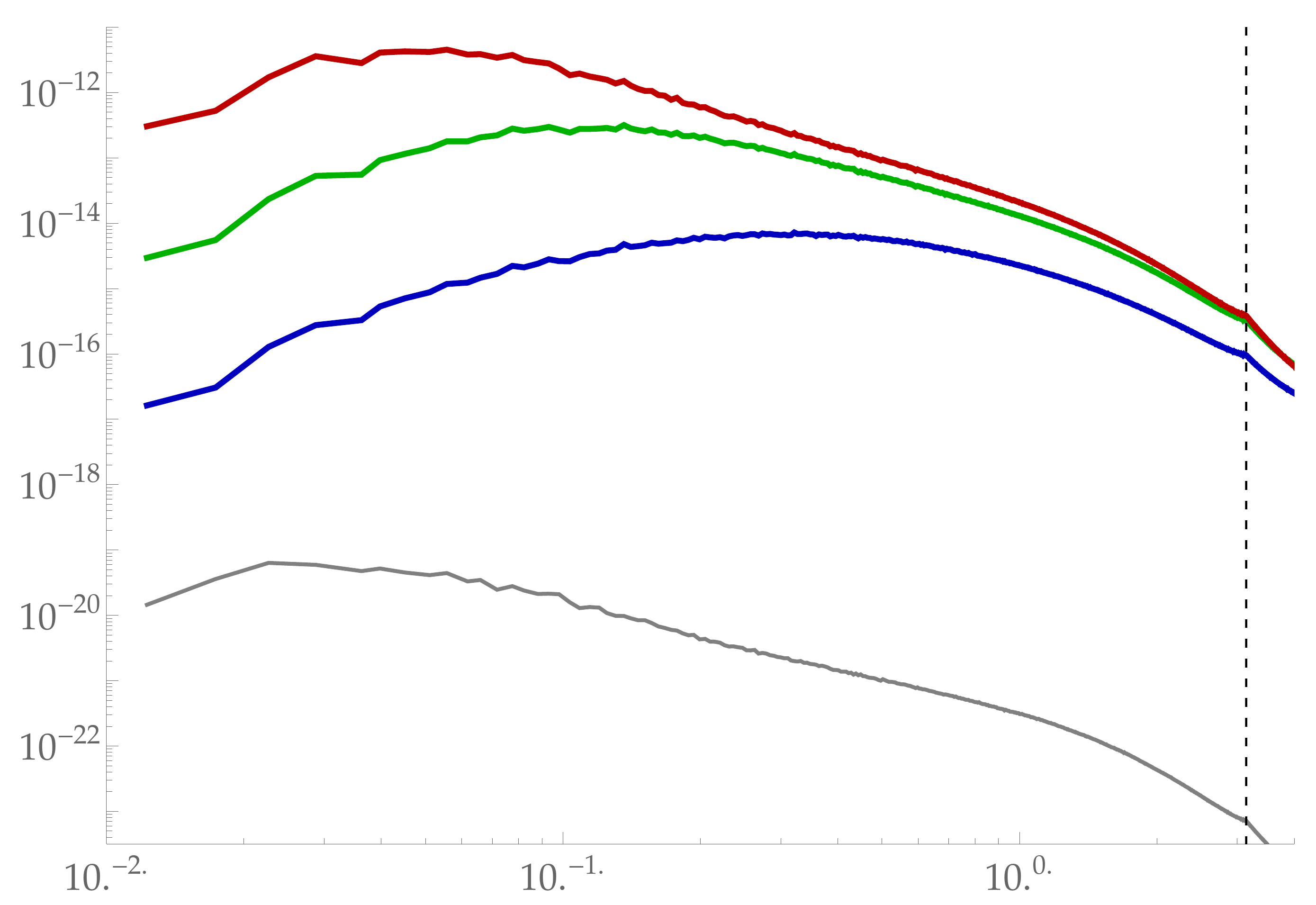}
      \fontsize{8}{0}\selectfont
      \put(46,-3)
      {
      $k$ [$h$/Mpc]
      }
      \put(-6,31)
      {
      \begin{sideways}
        $\mathcal P_\chi$
      \end{sideways}
      }
      \put(91,-1)
      {
      $ k_{\rm Ny}$
      }
    \end{overpic}
    \vspace*{.1em}
    \caption{Power spectra of $\chi$, compared with $\Lambda$CDM (grey line). As we can see, deviations can be enormous and remain large even at large scales and even for $|f_{R_0}| = 10^{-6}$, for which value the other quantities of interest essentially recover the $\Lambda$CDM solutions.}
    \label{fig_chi}
  \end{subfigure}
  \caption{Results for $\delta R$ and $\chi$.}
\end{figure}

We present our results for $\chi$ in figure~\ref{fig_chi}. In this case we show the actual power spectra instead of the relative power enhancement. As we have seen in \S\ref{sec_chi_and_xi}, the scalaron directly sources $\chi$ and so when $\delta f_R$ is much larger than $\chi$ would be in GR, it completely dominates and we have essentially $\chi \approx \delta f_R$. Because $\delta f_R$ can in principle be of the same order of magnitude as $|f_{R_0}|$ and even larger, we can easily see how $\chi$ can be many orders of magnitude larger than in $\Lambda$CDM.

Interestingly, this is the only quantity for which in the case $|f_{R_0}| = 10^{-6}$ we do not recover $\Lambda$CDM plus very small corrections, but instead deviations remain large, of several orders of magnitude, even at large scales. The possibility of detecting signatures of modified gravity using the gravitational slip is an interesting topic (see e.g. the recent~\cite{Pizzuti:2019wte}), and we plan on discussing some of these possibilities in an upcoming work~\cite{Reverberi_Davero_et_al_preparation}.

\section{Conclusions}
\label{sec_conclusions}

We have presented the framework and first results from the code \texttt{fRevolution}, based on the relativistic code \texttt{gevolution}~\cite{Adamek:2016zes}. We have discussed the approximation scheme which only relies on the weak field limit of GR with no further assumption on the smallness of density and scalar curvature perturbations, nor on the smallness of the scalaron $\delta f_R$ compared to $\bar f_R$. Moreover, we go beyond the Newtonian limit (see \S\ref{sec_newtonian_limit}) and take into account the Hubble friction when solving for the scalaron dynamics.

Overall, our results agree very well with analytical predictions in the case of the field produced by a point mass, and with existing (Newtonian) modified gravity codes for the matter power spectrum. To our knowledge, we present for the first time direct results for the scalar curvature perturbations, which however can be computed even in a strictly Newtonian framework, and for the gravitational slip $\chi$ and frame dragging $B_i$ which instead are intrinsically relativistic effects and can be computed in Newtonian codes only \text{a posteriori} under the assumption that Newtonian and relativistic solutions for $\Phi$ and $\delta f_R$ are essentially the same.

For the chosen parameter values, we detect a power excess for vector modes of about a factor 2, and differences of several orders of magnitude for $\chi$ and $\delta R$ compared to the $\Lambda$CDM predictions. While the impact of modified structure formation on the gravitational slip might provide interesting new directions to test and constrain modified gravity, the observed deviations from GR in the solutions for $\delta R$ suggest that extra care should be taken, when formulating predictions in modified gravity based on the assumption that deviations from the GR curvature are small, namely that not only $|f_R| \ll 1$, but also that $|\delta R + 8\pi G \delta T| \ll |8\pi G\delta T|$. We plan to carry out a more extensive discussion and analysis of our results in an upcoming publication~\cite{Reverberi_Davero_et_al_preparation}.

\begin{acknowledgments}
	The authors would like to thank J. Adamek, M. Kunz and I. Sawicki for useful comments and discussions during the development and testing of the code, and \'A. de la Cruz-Dombriz and P. Dunsby for their support and contribution during the early phases of the project. LR would like to thank the DAMTP, University of Cambridge for its hospitality. Early tests of the code have been carried out on the clusters \textit{Zeus} at the University of Cape Town and \textit{Baobab} at the University of Geneva. The final simulations have been carried out on \textit{Koios} at the Institute of Physics of the Czech Academy of Sciences in Prague. DD is supported by an Advanced Postdoc.Mobility grant of the Swiss National Science Foundation. LR is supported by ESIF and MEYS (Project CoGraDS-CZ.02.1.01/0.0/0.0/15\_003/0000437).
\end{acknowledgments}

\bibliography{./Biblio.bib}
\end{document}